\documentclass[12pt]{article}
\usepackage{bezier}
\usepackage{amssymb}
\usepackage{epsfig}

\newtheorem{lemma}{Lemma}[section]

\newtheorem{theorem}{Theorem}[section]

\newtheorem{definition}{Definition}[section]

\newcommand{\nc}{\newcommand}
\nc{\be}{\begin{equation}}
\nc{\ee}{\end{equation}}
\nc{\bea}{\begin{eqnarray}}
\nc{\eea}{\end{eqnarray}}
\nc{\bela}{\begin{eqnarray*}}
\nc{\eela}{\end{eqnarray*}}

\nc{\eqn}[1]{{(\ref{#1})}}
\nc{\cA}{{\cal A}}
\nc{\cB}{{\cal B}}
\nc{\cC}{{\cal C}}
\nc{\cD}{{\cal D}}
\nc{\cE}{{\cal E}}
\nc{\cF}{{\cal F}}
\nc{\cG}{{\cal G}}
\nc{\cH}{{\cal H}}
\nc{\cI}{{\cal I}}
\nc{\cJ}{{\cal J}}
\nc{\cK}{{\cal K}}
\nc{\cL}{{\cal L}}
\nc{\cM}{{\cal M}}
\nc{\cN}{{\cal N}}
\nc{\cO}{{\cal O}}
\nc{\cP}{{\cal P}}
\nc{\cQ}{{\cal Q}}
\nc{\cR}{{\cal R}}
\nc{\cS}{{\cal S}}
\nc{\cT}{{\cal T}}
\nc{\cU}{{\cal U}}
\nc{\cV}{{\cal V}}
\nc{\cW}{{\cal W}}
\nc{\cX}{{\cal X}}
\nc{\cY}{{\cal Y}}
\nc{\cZ}{{\cal Z}}

\nc{\simo}[1]{{\stackrel{#1}{\simeq}}}
\nc{\geqo}[1]{{\stackrel{#1}{\geq}}}
\nc{\geo}[1]{{\stackrel{#1}{>}}}
\nc{\guo}[1]{{\stackrel{#1}{\succ}}}

\nc{\rbo}{\raisebox}
\nc{\RR} {\rangle \! \rangle}
\nc{\LL} {\langle \! \langle}
\nc{\rmi}[1]{{\mbox{\small #1}}}
\nc{\eq}{eq.~}
\nc{\nr}[1]{(\ref{#1})}
\nc{\ul}{\underline}
\nc{\mc}{\multicolumn}
\nc{\todo}[1]{\par\noindent{\bf $\rightarrow$ #1}}

\nc{\cu}{{\cal u}}

\title{
  \begin{flushright} {\small $\begin{array}{ l } \mbox{HD--THEP--96--09} \\
    \mbox{MS--TPI--96--06} \end{array} $}
 \end{flushright}
\vskip 2cm
Linked Cluster Expansions
Beyond Nearest Neighbour Interactions:\\
Convergence and Graph Classes}

\author{Andreas~Pordt\thanks{\tt
       Email address pordt@poincare.uni-muenster.de}\\
        Institut f\"ur Theoretische Physik I, Universit\"at M\"unster, \\
        Wilhelm-Klemm-Str.\ 9, D-48149 M\"unster, Germany
        \and
        Thomas~Reisz\thanks{\tt Heisenberg fellow,
        Email address reisz@thphys.uni-heidelberg.de}
        \\Institut
        f\"ur Theoretische Physik,
        Universit\"at Heidelberg, \\
        Philosophenweg 16,
        D-69120 Heidelberg, Germany}

\begin{document}

\maketitle

\begin{abstract}
We generalize the technique of
linked cluster expansions on hypercubic lattices
to actions that couple fields at lattice
sites which are not nearest neighbours.
We show that in this case the graphical expansion can be arranged in such
a way that the classes of graphs to be considered are identical
to those of the pure nearest neighbour interaction.
The only change then concerns the computation of lattice
imbedding numbers.
All the complications that arise can be reduced to a generalization of
the notion of free random walks, including hopping beyond
nearest neighbour. Explicit expressions for combinatorical numbers
of the latter are given. We show that under some general conditions
the linked cluster expansion series have a non-vanishing
radius of convergence.
\end{abstract}


\section{Introduction}
Linked cluster expansions are series expansions about completely
disordered lattice systems \cite{Wortis,ID}.
Under very general conditions on the action the high temperature
series can be shown to have a non-vanishing radius of convergence
\cite{andreas1}.
In many cases the singularity closest to the origin can be related
to the physical singularity of the free energy, that is,
the phase transition.
To achieve valuable information from the series representations
one needs to compute high orders (for most recent
reference cf.~\cite{LW1,thomas1,butera,campost}).
This is best done by means of graph theoretical methods.
Usually the expansion is restricted to contact terms that couple
fields located at nearest neighbour lattice sites.
On the hypercubic lattice, which we refer to in the following,
this leads to a considerable reduction of the number of graphs.
The most striking property in this respect is that they have to have
an even number of lines in every loop.

Here we discuss the question of how to take care of interactions
beyond nearest neighbours
in an optimal way such that the classes of graphs
have to be modified minimally.
This implies to keep them reliably small in number.
As will be shown below, the graphs of nearest neighbour interactions
are sufficient for the general case also.
The only complication that arises can be absorbed completely
by a generalization of random walks, for which explicit
expressions are available.

Let $\Lambda$
denote a $D$-dimensional hypercubic lattice, either infinite
$\Lambda = {\mathbb{Z}}^D$ or
finite $\Lambda=\times_{i=0}^{D-1}({\mathbb{Z}}/L_i{\mathbb{Z}})$, with
all $L_i$ even (for convenience) and with periodic boundary
conditons imposed.
To be specific, here we discuss
models that are described by the partition function
\be \label{int.1}
   Z(J,v) = \int \prod_{x\in\Lambda} d^N\Phi(x) \;
   \exp{(-S(\Phi,v)+\sum_{x\in\Lambda}
   \sum_{a=1}^N J_a(x)\Phi_a(x) )},
\ee
where $\Phi$ denotes a real, $N$-component scalar field, and $J$ are
external sources.
The action is assumed to be of the form
\be \label{int.2}
   S(\Phi,v) =  \sum_x V(\Phi(x)) + \frac{1}{2}\;
   \sum_{x\not=y\in\Lambda}\sum_{a,b=1}^N \Phi_a(x)v_{ab}(x,y)\Phi_b(y),
\ee
with the lattice site action $V(\Phi)$, supposed to be
$O(N)$ invariant and appropriately bounded from below.

Truncated correlation functions are obtained by differentiation of
the generating functional
\bea
  W(J,v) & = & \ln{ Z(J,v) }, \nonumber \\
  \label{int.3}
  W_{a_1\ldots a_{2n}}^{(2n)} (x_1,\ldots, x_{2n} \vert v) & = &
   <\Phi_{a_1}(x_1) \cdots \Phi_{a_{2n}}(x_{2n}) >^c
   \\  & = & \left.
   \frac{\partial^{2n}}{\partial J_{a_1}(x_1) \cdots
   \partial J_{a_{2n}}(x_{2n})}
    W(J,v) \right\vert_{J=0}, \nonumber
\eea
Susceptibilities are defined as zero momentum correlations such as
\bea \label{int.4}
  \delta_{a,b}\; \chi_2 & = &
   \sum_x < \Phi_a(x)\Phi_b(0) >^c \\ \label{int.5}
  \delta_{a,b}\; \mu_2 & = & \sum_x (\sum_{i=0}^{D-1} x_i^2)
    \, < \Phi_a(x)\Phi_b(0) >^c .
\eea
The linked cluster expansion is the Taylor expansion
of the generating functional $W(J,v)$ with respect to
$v(x,y)$,
\be \label{int.6}
  W(J,v) = \left. \left( \exp{\sum_{x,y} \sum_{a,b} v_{ab}(x,y)
   \frac{\partial}{\partial \widehat v_{ab}(x,y)}} \right) W(J,\widehat v)
   \right\vert_{\widehat v =0}.
\ee
It generalized in the obvious way to connected correlation functions
according to \eqn{int.3}. Multiple derivatives of $W$ with respect to
$v(x,y)$ are managed by the identity \cite{Wortis}
\be \label{int.7}
  \frac{\partial W}{\partial v_{ab}(x,y)} = \frac{1}{2} \left(
   \frac{\partial^2 W}{\partial J_a(x) \partial J_b(y)} +
   \frac{\partial W}{\partial J_a(x)} \frac{\partial W}{\partial J_b(y)}
   \right) .
\ee
In most cases under consideration, the hopping term of the action
is assumed to couple fields only at nearest neighbour lattice sites,
that is,
\be \label{int.8}
   v_{ab}(x,y) \; = \; \left\{
   \begin{array}{r@{\qquad ,\quad} l }
    2\kappa\;\delta_{a,b}\; & {x=y\pm \widehat\mu \rm\; for\; some \; \mu,} \\
    0 & {\rm otherwise.}
   \end{array} \right.
\ee
$\widehat\mu$ denotes the unit vector in the $\mu$th direction,
$\mu=0,\ldots,D-1$. Correlation functions and susceptibilities
become series in the hopping parameter $\kappa$, with coefficients
depending on the toplogy of the lattice and the interaction $V$.
If $V$ satisfies appropriate bounds, the series can be shown to have
a non-zero radius of convergence, and convergence is uniform in volume
\cite{andreas1}.

In order to manage the complexities which increase rapidly with
increasing order, one is led in a natural way to a graphical device.
Truncated correlation functions get a representation as a sum over
connected graphs, each one endowed with the appropriate weight.
In this description a line of a graph represents a nearest neighbour
contact term. It implies the constraint that the two attached vertices
have to be placed at nearest neighbour lattice sites.
In this way, selflines are excluded, but otherwise no additional
exclusion constraints hold. This leads to considerable simplifications
and fast methods to compute weights of graphs.
In particular, on the hypercubic lattice,
for the weight of a graph to be non-vanishing,
every occuring loop in this graph is subject to the constraint that
it has an even number of lines. In turn, the number of graphs
can be reduced in a considerable way.
A graph with $L$ lines contributes to the order $\kappa^L$.

The simplications just mentioned do no more apply for more general
contact terms, that is, if $v$ couples fields at lattice sites
with larger separations.
A line still represents an interaction $v$ but otherwise no nearest
neighbour constraint holds anymore.
At the first sight this amounts to a considerable enlargement
of the class of graphs to be considered. In turn all the
algorithms developed for the construction of graphs to high orders
have to be extended as well.

In this letter we show that the arising complications do not have any
influence on the graphical expansions.
Hopping parameter expansions for lattice models with pair interactions
over larger distances can be cast into graph classes identical to those
already known from pure nearest neighbour interactions.
This implies that the same algorithms to generate all contributing
graphs apply to this case.
The only complication that arises is that the notion of
free random walks assigned to chains of vertices with two neighbours
have to be generalized appropriately.

In the second part of this paper we will generalize the proof given in
\cite{andreas1}, namely
that the radius of convergence is non-vanishing
for linked cluster expansions of models with nearest
neighbour interactions,
to the present case.
It will be shown that
under some general conditions the connected
Green functions and susceptibilities are analytic functions in
the hopping parameters for a large class of pair interactions.

%
%

\section{\label{basic.0}Graphical expansion.}

\subsection{Some basic graph theory}

A graph is a sequence of two objects and two maps
\[
  \Gamma = (\cL_\Gamma,\cB_\Gamma,E_\Gamma,\Phi_\Gamma),
\]
with $\cL_\Gamma$ and $\cB_\Gamma\not=\emptyset$ disjoint sets,
the internal lines and vertices of $\Gamma$, respectively.
$E_\Gamma$ is a map
\bea
  E_\Gamma: \cB_\Gamma & \to & \{0,1,2,\ldots\}, \nonumber \\ \label{lce.16}
   v & \to & E_\Gamma(v),
\eea
that assigns to every vertex $v$ the number of external lines
$E_\Gamma(v)$ attached to it.
The number of external lines of $\Gamma$ is given by
$E_\Gamma=\sum_{v\in\cB_\Gamma} E_\Gamma(v)$.
Finally, $\Phi_\Gamma$ is the incidence relation that assigns internal lines
to their endpoint vertices.
We consider lines as being unoriented, so $\Phi_\Gamma$
maps onto unoriented pairs of vertices
\be \label{lce.18}
  \Phi_\Gamma: \cL_\Gamma \to \overline{(\cB_\Gamma\times\cB_\Gamma)} .
\ee
In order to stay more general we do not exclude selflines in this
definition. External
vertices are those that have external lines attached,
\be \label{lce.19}
  \cB_{\Gamma,ext} =  \{ v\in\cB_\Gamma \; \vert \;
   E_\Gamma(v)\not=0 \},
\ee
whereas internal vertices don't,
$\cB_{\Gamma,int}=\cB_\Gamma\setminus\cB_{\Gamma,ext}$.
For every pair of vertices $v,w\in\cB_\Gamma$,
$m(v,w)$ denotes the number of common lines of $v$ and $w$,
i.e.~the number of elements of
\be \label{lce.20}
  \cM(v,w) = \{ l\in\cL_\Gamma \; \vert \;
   \Phi_\Gamma(l)= \overline{(v,w)} \},
\ee
\[
   m(v,w) \; = \; | \cM(v,w) |.
\]
$w$ is called a neighbour of $v$ if $m(v,w)\not=0$, and
\be \label{lce.21}
  \cN(v) = \{ w\in\cB_\Gamma \; \vert \; m(v,w)\not=0 \}
\ee
is the set of neighbours of $v$, with $N(v)=|\cN(v)|$
its number of elements.
For this definition, multiple lines are not count according to
their multiplicity. For every integer $n$,
$n$-vertices are those that have precisely $n$ neighbours,
\be \label{lce.22}
  \cB_\Gamma^{(n)} = \{ v\in\cB_\Gamma \; \vert \;
   N(v)=n \}.
\ee
An $n$-vertex has at least $n$ internal lines attached.
The number of total lines, both internal and external ones,
attached to a vertex $v$ is denoted by
\[
  l(v) := \sum_{w\in\cB_\Gamma} m(v,w) + E_\Gamma(v) .
\]
If $l(v)$ is even for every $v\in\cB_\Gamma$, $\Gamma$ is called
even. In particular, then,
$\Gamma$ has an even number of external lines.
A vertex $v\in\cB_\Gamma$ is called weak $n$-vertex,
$v\in\cB^{(n)}_{\Gamma,\rm weak}$, if
$N(v)=l(v)=n$. A weak vertex does not have any external line attached.

The following topological notions will be used below.
A path is an ordered non-empty sequence $p=(w_1,\ldots,w_n)$ of
vertices with $m(w_i,w_{i+1})\not=0$ for all $i=1,\ldots ,n-1$. $p$ is
called a path in $\Gamma$ from $w_1$ to $w_n$ of length $n-1$.
If in addition $w_n=w_1$, $p$ is called a loop, with $n-1$ its
number of lines.
If for every pair of vertices $v,w\in\cB_\Gamma$, $v\not=w$,
there is a path from $v$ to $w$, the graph $\Gamma$ is called connected.
A path $p=(w_1,w_2,\ldots ,w_n)$ is called a maximal weak 2-chain
if $w_i\in\cB_{\Gamma,\rm weak}^{(2)}$ for all $i=2,\ldots ,n-1$,
but $w_1,w_n\not\in\cB_{\Gamma,\rm weak}^{(2)}$.

\begin{figure}[htb]

\begin{center}
\setlength{\unitlength}{0.8cm}
\begin{picture}(10.0,4.0)


\put(2.0,0.30){
\setlength{\unitlength}{0.8cm}
\begin{picture}(10.0,3.0)

{\linethickness{0.2pt}
\qbezier(0.0,0.0)(2.0,2.0)(4.0,2.0)
}

\put(0.0,0.0){\circle*{0.15}}
\put(4.0,2.0){\circle*{0.15}}
\put(1.0,0.88){\circle*{0.15}}
\put(2.0,1.5){\circle*{0.15}}
\put(3.0,1.87){\circle*{0.15}}

\put(0.0,0.0){\line(-2,-1){0.8}}
\put(0.0,0.0){\line(-1,-2){0.4}}
\put(0.0,0.0){\line(0,-1){1.0}}

\put(4.0,2.0){\line(1,-1){0.5}}
\put(4.0,2.0){\line(1,1){0.5}}

\put(0.0,0.0){\makebox(1.0,0){$w_1$}}
\put(4.0,2.0){\makebox(1.0,0){$w_5$}}

\end{picture}
}

\end{picture}
\end{center}

\caption{\label{weakchain}
Example of a maximal weak 2-chain of length 4.
The 2-chain has 3 vertices belonging to $\cB_{\Gamma,\rm weak}^{(2)}$
and hence 4 lines.
}
\end{figure}
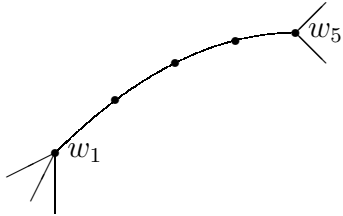


Two graphs
\be \label{lce.23}
  \Gamma_1 = (\cL_1,\cB_1,E_1,\Phi_1), \;
  \Gamma_2 = (\cL_2,\cB_2,E_2,\Phi_2)
\ee
are called (topologically) equivalent if there are two maps
$\phi_1:\cB_1\to \cB_2$ and $\phi_2:\cL_1\to \cL_2$,
such that
\bea \label{lce.24}
  \Phi_2 \circ \phi_2 &=& \overline\phi_1 \circ \Phi_1, \nonumber \\
  E_2 \circ \phi_1 &=& E_1,
\eea
where $\circ$ means decomposition of maps, and
\bea
  \overline\phi_1: \overline{\cB_1\times\cB_1} & \to &
      \overline{\cB_2\times\cB_2} \nonumber \\ \label{lce.24.1}
   \overline\phi_1(v,w)  & = & (\phi_1(v),\phi_1(w)).
\eea
A symmetry of a graph
$\Gamma = (\cL,\cB,E,\Phi)$ is a pair of maps
$\phi_1:\cB\to\cB$ and $\phi_2:\cL\to \cL$,
such that
\bea \label{lce.25}
  \Phi \circ \phi_2 &=& \overline\phi_1 \circ \Phi, \nonumber \\
  E \circ \phi_1 &=& E.
\eea
The number of those maps is called the symmetry number of $\Gamma$.

The set of equivalence classes of connected graphs with $E$ external
and $L$ internal lines is henceforth denoted by
$\cG_E(L)$, and
\be \label{lce.28}
  \cG_E \; := \; \bigcup\limits_{L\geq 0} \; \cG_E(L).
\ee
If we impose the additional constraint that every
loop has to have an even number of lines,
we get the sets $\cG_E^{\rm ev}(L)$ and $\cG_E^{\rm ev}$.

In practice one introduces further classes such as one-particle
(or better to say: one-line) irreducible ones.
Because this is of no importance for the following, we do not introduce
them here. Also, just for simplicity we discuss the one-component model only.
The generalizations are straightforward.

\subsection{Weights}

The graphical
representation of the linked cluster expansion of susceptibilities
is an expansion in term of equivalence classes of connected graphs as
defined above. For every such equivalence class we need exactly one
representative.

Every graph $\Gamma$ represents a number, which is called its weight.
If $\Gamma$ is connected and has $E$ external lines, it contributes this
weight to the susceptibility $\chi_E$. For the one-component field models,
it is computed along the following lines.
Only even graphs need to be considered, otherwise the weight vanishes.

\begin{description}

\item{1.} All the vertices of $\Gamma$ are placed at lattice sites.
No exclusion principle holds, i.e.~any number of vertices can be placed
at the same site. By assumption, the model lives on the lattice
$\Lambda$, which satisfies translation invariance. An arbitrary selected
vertex is located at a fixed lattice site, avoiding a volume factor.
A priori, all the other vertices of the graph are placed arbitrarily.

\item{2.} Every internal line of $\Gamma$ is assigned a factor $v(x,y)$,
where $x,y\in\Lambda$ are the lattice
sites its endpoints are placed at.

\item{3.} The sum is taken over all possible placements of the vertices at
lattice sites (except for the fixed vertex).
The amount of computational work can be reduced considerably by taking
into account the particular form of $v(x,y)$ from the very beginning.

\item{4.} A factor $\stackrel{\circ}{v}_{2n}^c$
for every vertex $v\in\cB_\Gamma$, where
\be \label{weight.1}
   \stackrel{\circ}{v}_{2n}^c =
   W^{(2n)} (0,\ldots, 0 \vert v=0),
\ee
cf.~Eqn. \eqn{int.3}, and $2n$ is the sum of internal and external
lines attached to $v$.

\item{5.} Two final factors complete the weight. The first one is given by
$(S(\Gamma))^{-1}$, where $S(\Gamma)$ is the symmetry number of $\Gamma$.
It has the product representation
\be \label{lce.26}
   S(\Gamma) = S_P(\Gamma) \cdot
   \prod_{(v,w)\in\overline{\cB_\Gamma\times\cB_\Gamma}} m(v,w)!,
\ee
where the integer number $S_P(\Gamma)$ corresponds to the symmetries
of $\Gamma$ under
permutations of vertices. The latter product is due to the exchange of
multiple lines between vertices.
The other final factor counts the various enumerations of the
external lines,
\be \label{lce.27}
    \frac{ E! }{\prod_{v\in\cB_\Gamma} E_\Gamma(v)!}.
\ee

\end{description}

Weighted susceptibilities such as the moment $\mu_2$
are described by the same diagrams as unweighted ones. The only
difference is that they have the appropriate weight factors inserted.
This, of course, must be done before the sum over the lattice
imbeddings is carried out.

%
%

\section{\label{res.0}Reorganization of graphs.}
Let us repeat that $\Lambda$
denotes the $D$-dimensional hypercubic lattice
$\Lambda=\times_{i=0}^{D-1}{\mathbb{Z}}/L_i{\mathbb{Z}}$, with
all $L_i$ even, possibly infinity, and with periodic boundary
conditions imposed.
In terms of the classes of graphs introduced in the last section,
truncated correlation functions and in particular susceptibilities
allow for a representation as sum over graphs
\be \label{res.1}
  \chi_E \; = \; \sum_{\Gamma\in\cG_E} w(\Gamma),
\ee
with corresponding weight $w(\Gamma)$.
At this point we are not concerned with questions of convergence.
Our interest here is in combinatorical rearrangements to arbitrary
large but finite order.
If desired we limit the discussion to propagators with
$v(x,y)=0$ if $||x-y||>R$ for convenient $R>0$,
where
\be \label{res.2}
  ||x-y|| \; = \; \sum_{i=0}^{D-1}
  \inf_{n\in\mathbb{Z}} |x_i-y_i+n L_i|,
\ee
and to graphs $\Gamma\in\cG_E$ with a limited number of lines.
The limitation is not necessary for a convergence proof of
the linked cluster expansion. This will be discussed in the last
section of this work.
We also mention in passing that the discussion below works also
for more restricted classes of graphs such as
1-vertex irreducible or 1-particle irreducible ones.

In the following we restrict attention to pair interactions
that respect the euclidean lattice symmetries.
A convenient notion in this respect is given by

\begin{definition}\label{resdef.1}
A signature $p$ of length $l\in\mathbb{N}$ is a collection of $D$
non-negative integers $l_i\in \mathbb{N}\cup \{ 0\} $,
\be \label{res.3}
p=(l_0,\ldots ,l_{D-1}),
\ee
satisfying $\sum_{i=0}^{D-1} l_i =l$.

Collections of integers are unordered sequences, i.e. permutations
of elements in $p$ represent the same signature.
For a signature $p$ we write $l_p$ for its length,
and the set of ordered D-tupels originating from $p$
is denoted by $\cP_D(p)$.
\end{definition}

We may represent a signature $p=(l_0,\ldots ,l_{D-1})$ as a path
connecting two sites $x,y$ on the $D$-dimensional lattice. The
path is given by $\{ (xx_1),(x_1x_2),\ldots ,(x_{D-1}y)\} ,$
$x_{i+1} = x_i + l_i e_i \widehat i ,$ $i\in \{ 0,\ldots ,D-1\} ,$
$e_i \in \{ -1,1\} ,$
$x_0 = x,$ $x_{D} = y$, where $\widehat i$ is the unit vector in
$i$-direction.

We consider hopping propagators of the form
\be \label{res.6}
   v(x,y) \; = \; \sum_{p\in\cS} 2\kappa_p \; v^p(x,y),
\ee
where $\cS$ is a finite set of signatures $p$ of lengths $l_p\geq 1$,
and with
\be \label{res.7}
   v^p(x,y) \; = \; \sum_{l\in\cP_D(p)} \;
   \sum_{e_0,\ldots ,e_{D-1}=\pm 1}
   \delta^{\rm T}_{x-y,\sum_{i=0}^{D-1} l_ie_i \widehat i} \; .
\ee
$\delta^{\rm T}$ denotes the Kronecker delta on the torus,
that is,
\[
   \delta^{\rm T}_{x,y} = 1 \quad \mbox{if and only if} \quad
   \inf_{n\in\mathbb{Z}} | x_j-y_j+n L_j| = 0 \;\mbox{for all}\;
   j = 0,\ldots ,D-1,
\]
and $\delta^{\rm T}_{x,y}=0$ in all other cases.
Eqn.~\eqn{res.7} denotes a propagator of a particular geometric profile.
We assume that every $p\in\cS$ of the form \eqn{res.3}
satisfies for all $j=0,1,\ldots ,D-1$
\[
   l_j \; < \min_{i=0,\ldots,{D-1}} \; \frac{L_i}{2}.
\]
In particular, then, we have for $p\in\cS$ of length $l_p$
\be \label{res.zero}
   v^p(x,y) = 0 \; \mbox{if} \; ||x-y|| \not= l_p.
\ee
The assumption made here that $\cS$ is a finite set is
just for convenience.
Sufficient convergence conditions that allow $\cS$ to be infinite
are discussed in the next section.

Our aim is to cast \eqn{res.1} in such a form that we have to sum
over graphs only that we have met already for the case of pure
nearest neighbour interactions. That is,
\be \label{res.8}
  \chi_E \; = \; \sum_{\Gamma\in\cG_E^{\rm ev}} w^{\rm ev}(\Gamma),
\ee
with appropriate weight $w^{\rm ev}(\Gamma)$.
Toward this end, let $\widetilde\cG_E$ denote the set of pairwise
inequivalent graphs with $E$ external lines and no weak 2-vertices,
that is,
\be \label{res.9}
  \widetilde\cG_E \; = \;
  \left\{ \Gamma\in\cG_E \vert
    \cB_{\Gamma,{\rm weak}}^{(2)}  = \emptyset \right\}.
\ee
We obtain from \eqn{res.1}
\be \label{res.10}
  \chi_E \; = \; \sum_{\Gamma\in\widetilde\cG_E} \widetilde{w}(\Gamma),
\ee
where the weight is computed as outlined in the last section,
the only modification being that a line now represents,
instead of $v(z,w)$, a factor
\bea
   P_{z\to w}( (\kappa_p)_{p\in\cS}, \stackrel{\circ}{v}_2^c)
   & = & \sum_{m\geq 1} \; \sum_{x_1,\ldots ,x_{m-1}\in\Lambda}
   v(z,x_1) v(x_1,x_2) \cdots
   v(x_{m-1},w) \; (\stackrel{\circ}{v}_2^c)^{m-1} \nonumber \\
   \label{res.11}
   & = & \frac{1}{\stackrel{\circ}{v}_2^c}
   \left( 1 - \stackrel{\circ}{v}_2^c v\right)^{-1}(z,w).
\eea
The next step is to decompose $P_{z\to w}$ in terms of random walks
of given length in units of lattice links. This is formulated as

\begin{lemma}\label{reslemma.1}
For the hopping propagator $v$ given by \eqn{res.6}, we have
\be \label{res.12}
    P_{z\to w}( (\kappa_p)_{p\in\cS}, \stackrel{\circ}{v}_2^c) =
    \sum_{l\geq 1} P_{z\to w}^l
    ( (\kappa_p)_{p\in\cS}, \stackrel{\circ}{v}_2^c),
\ee
with
\bea
   P_{z\to w}^l ( (\kappa_p)_{p\in\cS}, \stackrel{\circ}{v}_2^c)
   &=& \sum_{m=1}^l \left( \stackrel{\circ}{v}_2^c \right)^{m-1}
   \sum_{ p_1,\ldots ,p_m \in\cS \atop (\sum_{i=1}^m l_{p_i}=l) }
   \quad \sum_{x_1,\ldots ,x_{m-1}\in\Lambda} \nonumber \\
   && \label{res.13} \\
   && \qquad \cdot v^{p_1}(z,x_1) \cdots
   v^{p_m}(x_{m-1},w) \; (2\kappa_{p_1})\cdots(2\kappa_{p_m}) \nonumber \\
   && \nonumber \\
   \label{res.14}
   &=& \sum_{ (n_p|p\in\cS) \atop (l=\sum_{p\in\cS} n_p l_p) }
   \left( \stackrel{\circ}{v}_2^c \right)^{\sum_{p\in\cS} n_p-1}
   \cN_{z\to w}^{(n_p|p\in\cS)}
   \prod_{p\in\cS} (2\kappa_p)^{n_p},
\eea
where we have set $x_0=z$, and with non-negative integers
\be \label{res.15}
   \cN_{z\to w}^{(n_p|p\in\cS)} \; = \;
   \left( \prod_{p\in\cS} \frac{1}{n_p!}
   \frac{\partial^{n_p}}{\partial\kappa_p^{n_p}} \right) \cdot
   \left. \left( 1 - \sum_{p\in\cS} \kappa_p v^p \right)^{-1}(z,w)
   \right|_{\kappa_p\equiv 0}.
\ee
Furthermore
\be \label{res.16}
   P_{z\to w}^l ( (\kappa_p)_{p\in\cS}, \stackrel{\circ}{v}_2^c)
   \not= 0 \quad\mbox{{\rm only if}}\quad
   l - || z-w || \; \geq \; 0
   \quad \mbox{{\rm and even}}.
\ee
\end{lemma}

{\sl Proof:}
Inserting \eqn{res.6} into the right hand side of the first equality
of \eqn{res.11} and collecting the sum over all products of
propagators $v^p$ of fixed total length
$l=\sum_i l_{p_i}$, we obtain \eqn{res.12}, with $P^l_{z\to w}$
as given by \eqn{res.13}.
Furthermore, \eqn{res.14} follows immediately from the Taylor
expansion of
\[
   P_{z\to w}( (\kappa_p)_{p\in\cS}, \stackrel{\circ}{v}_2^c)
   \; = \; \frac{1}{\stackrel{\circ}{v}_2^c}
   \left( 1 - \stackrel{\circ}{v}_2^c \sum_{p\in\cS}
   (2\kappa_p) v^p \right)^{-1}(z,w).
\]
Finally, for every term of \eqn{res.13} we have,
with $x_0=z$ and $x_m=w$
\[
     || z - w ||  =  || \sum_{i=1}^{m-1} ( x_i - x_{i-1} ) ||
     \leq \sum_{i=1}^{m-1}  || x_i - x_{i-1} ||
      = \sum_{i=1}^{m-1} l_{p_i} = l ,
\]
and we have used \eqn{res.zero} and the triangle inequality
for the norm \eqn{res.2}. For $j=0,\ldots ,D-1$,
\bea
  \inf_{n\in\mathbb{Z}} | z_j-w_j + n L_j |
  & = & \inf_{n\in\mathbb{Z}} | \sum_{i=1}^{m-1} ( x_i-x_{i-1} )_j
   + n L_j | \nonumber \\
  & = & \sum_{i=1}^{m-1} \left(
   \inf_{n\in\mathbb{Z}} | ( x_i-x_{i-1} )_j + n L_j | \right)
   + \; \mbox{even number}. \nonumber
\eea
Hence,
\bea
     || z - w || & = & \sum_{j=0}^{D-1}
     \inf_{n\in\mathbb{Z}} | z_j - w_j + n L_j | \nonumber \\
   & = & \sum_{i=1}^{m-1} || x_i - x_{i-1} || +
   \; \mbox{even number} \nonumber \\
   & = & l + \; \mbox{even number}, \nonumber
\eea
that is, \eqn{res.16}.
This completes the proof.
$\qquad\square$

Inserting the decomposition \eqn{res.12} into
\eqn{res.10}, we obtain the desired representation \eqn{res.8}.
The property \eqn{res.16} ensures that only those graphs contribute
which have an even number of lines in each of their loops.

Hence we end up with the same class of graphs as for the case of
pure nearest neighbour interactions.
The weight of a graph is computed as described in the last section
with the following exception only.
Every maximal weak 2-chain of a graph contributes a factor
$P_{z\to w}^l ( (\kappa_p), \stackrel{\circ}{v}_2^c)$
to the lattice imbedding number,
where $l$ is the length of the chain and $w,z$ the lattice
sites where its two endpoint vertices are placed at. We notice that
for even $l$
neither these endpoint vertices have to be different nor the lattice
sites $z$ and $w$.
If $l=1$, $P_{z\to w}^l$ just imposes the nearest neighbour constraint
on $z$ and $w$.

If we choose the hopping parameters $(\kappa_p |p\in\cS)$ according to
\[
   \kappa_p \; = \; c_p \; \kappa^{l_p},
\]
for all signatures $p\in\cS$, with the $c_p$ being fixed,
every graph $\Gamma\in\cG_E^{\rm ev}$ with $L$ lines that contributes to
\eqn{res.8} has a weight proportional to $\kappa^L$.
In this case, every graph contributes to the same order of $\kappa$
as it would for the pure nearest neighbour interaction.

Translation invariance amounts to Fourier transform. With
\be \label{res.int}
   \int_k \; := \; \prod_{i=0}^{D-1} ( \frac{1}{L_i} \;
   \sum_{k_i = \frac{2\pi}{L_i} \nu_i \atop
    -\frac{L_i}{2} < \nu_i \leq \frac{L_i}{2} } )
\ee
we have
\[
   \sum_{e_0,\ldots ,e_{D-1} = \pm 1}
   \delta^{\rm T}_{x, \sum_i e_i n_i \widehat i } \; = \;
   \int_k {\rm e}^{i k\cdot x} \;
   \prod_{i=0}^{D-1} \left( 2 \cos{k_i n_i} \right).
\]
The random walk numbers $\cN_{z\to w}^{(n_p|p\in\cS)}$
are then given by
\be \label{res.fourier}
    \cN_{z\to w}^{(n_p|p\in\cS)}  =
    \frac{ (\sum_p n_p )! }{ \prod_p n_p! } \;
    \int_k {\rm e}^{i k\cdot (z-w)} \;
    \prod_{p\in\cS} \left( \sum_{l\in\cP_D(p)} \prod_{i=0}^{D-1}
    ( 2 \cos{k_i l_i} ) \right)^{n_p} .
\ee
Similar as for the pure nearest neighbour interactions,
this representation can be used to compute the random walk
combinatorical integers  $\cN_{z\to w}^{(n_p|p\in\cS)}$
explicitly.

As an illustrative example we supplement the
nearest neighbour couplings
by next-to-nearest neighbour interactions.
In this case, we have 3 signatures, $\cS=\{ p_1,p_2,p_3 \}$,
with


\begin{center}
\setlength{\unitlength}{0.8cm}
\begin{picture}(10.0,3.0)


\put(0.0,0.0){
\setlength{\unitlength}{0.8cm}
\begin{picture}(3.0,5.0)
\put(0.0,0.0){\circle*{0.17}}
\put(1.0,0.0){\circle*{0.17}}
\put(0.0,0.0){\line(1,0){1.0}}
\put(0.0,2.5){\makebox(1.0,0.0){$p_1=(1,0,0)$}}
\end{picture}
}


\put(4.0,0.0){
\setlength{\unitlength}{0.8cm}
\begin{picture}(3.0,5.0)
\put(0.0,0.0){\circle*{0.17}}
\put(1.0,1.0){\circle*{0.17}}
\put(0.0,0.0){\line(1,0){1.0}}
\put(1.0,0.0){\line(0,1){1.0}}
\put(0.0,2.5){\makebox(1.0,0.0){$p_2=(1,1,0)$}}
\end{picture}
}


\put(8.0,0.0){
\setlength{\unitlength}{0.8cm}
\begin{picture}(3.0,5.0)
\put(0.0,0.0){\circle*{0.17}}
\put(1.0,0.0){\circle{0.17}}
\put(2.0,0.0){\circle*{0.17}}
\put(0.0,0.0){\line(1,0){0.83}}
\put(1.17,0.0){\line(1,0){0.83}}
\put(0.0,2.5){\makebox(1.0,0.0){$p_3=(2,0,0)$}}
\end{picture}
}

\end{picture}
\end{center}


\noindent
written here for $D=3$ dimensions.
A maximal weak 2-chain of length $l$ as part of a graph
implies a factor
$P_{z\to w}^l ( (\kappa_p)_{p\in\cS}, \stackrel{\circ}{v}_2^c)$,
cf.~\eqn{res.13},\eqn{res.14},
with $z,w$ the lattices sites its end vertices are placed at.
In order to get lattice imbedding numbers,
one needs to compute the random walk numbers
$\cN_{z\to w}^{(n_1,n_2,n_3)}$
for all nonnegative integers $n_1,n_2,n_3$ with
\[
  l = n_1 + 2 ( n_2 + n_3 ).
\]
In Table \ref{tab1} we give all the numbers for $z=w$ and chain length
$l=16$.


\begin{table}[htb]
\caption{\label{tab1} Example of next-to-nearest neighbour
combinatorical random walk numbers
$\cN_{z\to w}^{(n_1,n_2,n_3)}$ on the
3-dimensional hypercubic lattice,
infinitely extended in all directions. The numbers are for $z=w$ and
total chain length $l=16$.
The pure nearest neighbour case gives
$\cN_{0\to 0}^{(16,0,0)}= 27770358330$.
}
\vspace{0.5cm}

\begin{center}

\begin{tabular}{|c|c|c|r||c|c|c|r|}
\hline \hline
$n_1$ & $n_2$ & $n_3$ & $\cN_{0\to 0}^{(n_1,n_2,n_3)}\quad$ &
$n_1$ & $n_2$ & $n_3$ & $\cN_{0\to 0}^{(n_1,n_2,n_3)}\quad$ \\
[0.5ex] \hline
14 &  1 &  0 &  137889591840  &
 4 &  4 &  2 &  20325816000 \\

14 &  0 &  1 &  56502916470  &
 4 &  3 &  3 &  11722233600  \\

12 &  2 &  0 &  277007722992 &
 4 &  2 &  4 &  3890527200  \\

12 &  1 &  1 &  227358907776 &
 4 &  1 &  5 &  678585600  \\

12 &  0 &  2 &  48295573326  &
 4 &  0 &  6 &  54299700  \\

10 &  3 &  0 &  288364684608 &
 2 &  7 &  0 &  437724000  \\

10 &  2 &  1 &  355550656032  &
 2 &  6 &  1 &  1266904800  \\

10 &  1 &  2 &  151142397120  &
 2 &  5 &  2 &  1617174720  \\

10 &  0 &  3 &  21791715660  &
 2 &  4 &  3 &  1175005440  \\

 8 &  4 &  0 &  165353348160  &
 2 &  3 &  4 &  509120640  \\

 8 &  3 &  1 &  272212617600 &
 2 &  2 &  5 &  142067520  \\

 8 &  2 &  2 &  173721477600  &
 2 &  1 &  6 &  18506880  \\

 8 &  1 &  3 &  50101286400  &
 2 &  0 &  7 &  1610280  \\

 8 &  0 &  4 &  5517544230  &
 0 &  8 &  0 &  4038300  \\

 6 &  5 &  0 &  51139740960 &
 0 &  7 &  1 &  13305600  \\

 6 &  4 &  1 &  105386339520  &
 0 &  6 &  2 &  20049120  \\

 6 &  3 &  2 &  89715669120 &
 0 &  5 &  3 &  17015040  \\

 6 &  2 &  3 &  38870314560  &
 0 &  4 &  4 &  9918720  \\

 6 &  1 &  4 &  8536207680  &
 0 &  3 &  5 &  2983680  \\

 6 &  0 &  5 &  761454540 &
 0 &  2 &  6 &  1028160  \\

 4 &  6 &  0 &  7704774000  &
 0 &  1 &  7 &  0 \\

 4 &  5 &  1 &  19072972800  &
 0 &  0 &  8 &  44730  \\ \hline

\end{tabular}

\end{center}

\end{table}


%
%

\section{\label{con.0}Convergence of the linked cluster expansion}
It was shown in \cite{andreas1} that the linked cluster expansion for
connected Green functions and susceptibilities have a non-zero
radius of convergence for the case of nearest neighbour couplings,
i.e.~$v^p =0$ for all signatures with length $l_p >1.$
We will generalize the proof given in \cite{andreas1} to hopping propagators
which contain terms beyond nearest neighbour interactions.

We do not suppose here that there are only a finite number of
terms in the hopping propagator eq. (\ref{res.6}). The non-local hopping
parameter may connect arbitrary lattice sites. $\cS $ is the set of all
signatures $p$ of length greater or equal to one.

Let us define a hopping parameter $\kappa $ and constants $c_p$ for all
signatures $p$ by
\[
\kappa_p = c_p \kappa^{l_p}.
\]
$l_p$ denotes the length of signature $p.$
The constants $c_p$ have to be chosen such that the series on the
right hand side of
eq.~(\ref{res.6}) is convergent. The next lemma gives a sufficient
condition on the coefficients $c_p$.
\begin{lemma} \label{lem1}
Suppose that there exists a positive constant $\overline\kappa >0$ such that
for all signatures $p$
\begin{equation} \label{e2}
|c_p| \le (2^{-1-\eta} \overline\kappa^{-1})^{l_p} \; \overline\kappa ,
\end{equation}
where $\eta >0$ is a constant obeying
\begin{equation} \label{e3}
\sum_{y:\, 0\not=y\in \Lambda} 2^{-\eta \Vert y\Vert } \le 4D \, 2^{-D} .
\end{equation}
Then a bound for the non-local part of the action is given by
\begin{equation} \label{e4}
|\frac{1}{2} \sum_{x,y\in \Lambda} \Phi (x) v(x,y) \Phi (y) | \le
 4D |\kappa | \sum_{x\in \Lambda } |\Phi (x) |^2
\end{equation}
for all complex $\kappa $, $|\kappa | \le \overline\kappa $ and
all $\Phi \in\cH (\Lambda )$
(=Hilbert space of square summable functions on
$\Lambda $).
\end{lemma}
\par\noindent
{\em Proof :} The number of all signatures $p$ with length $l_p=l$ is
bounded by
\begin{equation} \label{e5}
|\{ \mbox{$p$ signature}|\, l_p = l \} | \le 2^{D+l} .
\end{equation}
Then the left hand side of inequality (\ref{e4}) is bounded by
\[
  \sum_{x,y\in \Lambda} \sum_{p:\atop l_p = \Vert x-y\Vert}
 |c_p | |\kappa |^{l_p} |\Phi (x)|^2.
\]
Using supposition (\ref{e2}) the bound becomes
\begin{equation} \label{e999}
  {|\kappa |} \sum_{x,y\in \Lambda}
\sum_{p:\atop l_p = \Vert x-y\Vert} 2^{-l_p}\, 2^{-\eta \Vert x-y\Vert } \,
 |\Phi (x)|^2 .
\end{equation}
By the bounds (\ref{e5}) and (\ref{e999})
the left hand side of inequality (\ref{e4}) is estimated by
\[
 {2^D|\kappa |} \sum_{x,y\in \Lambda \atop x\not= y}
 2^{-\eta \Vert x-y\Vert } \,
 |\Phi (x)|^2 .
\]
Using supposition (\ref{e3}) we obtain the assertion.
$\qquad \square $

The convergence of the linked cluster expansion is presented in the following
theorem for a symmetric translation invariant
scalar field theory on an infinite lattice. A modification
of the proof for $N$-component models or models on a torus is
straightforward. Also the restriction to translation invariance and
$\mathbb{Z}_2$-symmetry is of no relevance for the convergence proof.
\begin{theorem} \label{the1}
Let a translation invariant and $\mathbb{Z}_2$-symmetric model be defined
on the lattice $\Lambda = \mathbb{Z}^D$
by  the partition function in eq. (\ref{int.1}). Suppose that the
supposition of lemma \ref{lem1} is valid and there
exist a positive $c > 0$ and a real $\delta $
such that for all $\Phi \in \mathbb{R}$
\begin{equation} \label{Stab}
V(\Phi ) \ge c\Phi^2 - \delta.
\end{equation}
Then there exists a positive constant $\kappa_*$ such that
for all $x,y\in \Lambda $
the 2-point Green func\-ti\-on $W^{(2)}(x,y)$ is an analytic function
in $K(\kappa_*) := \{ \kappa \in \mathbb{C} |\, |\kappa | < \kappa_* \}$.
Furthermore, there exist a constant $\alpha>0$
and for all $\kappa \in K(\kappa_*)$ there is $m(\kappa)>0$ such that
for all $x,y\in \Lambda $
\begin{equation} \label{Gbound}
|W^{(2)}(x,y)| \le  e^\alpha \, q_V^{(2)}\delta_{xy} +
\alpha \, \exp \{ -m\Vert x -y\Vert \},
\end{equation}
where
\begin{equation} \label{q2def}
q_V^{(2)} := \frac{\int d\Phi \, \Phi^2 \,
  \exp \{ -V(\Phi )\}}{\int d^N\Phi \, \exp \{ -V(\Phi )\}} .
\end{equation}
Furthermore, for $\kappa \rightarrow 0$,
\begin{equation} \label{Mass}
m(\kappa ) = O(\ln |\kappa |) > 0.
\end{equation}
\end{theorem}
For $m>0$ the series in the definition (\ref{int.4}) of the
2-point susceptibility $\chi_2 $ is convergent.
Thus, using a well-known theorem for analytic functions,
the above theorem implies that the 2-point susceptibility $\chi_2 $
exists and is an analytic function for $\kappa$ in
$K(\kappa_*) := \{ \kappa \in \cC |\, |\kappa | < \kappa_* \} $
for some $\kappa_*>0$.

Define a positive number
\begin{equation} \label{qdef}
q_V := \int d\Phi \exp \{ -V(\Phi )\}
\end{equation}
and for all finite subsets $X$ of $\Lambda $ the partition function
\begin{equation} \label{e7}
Z(X|J,\kappa ) = q_V^{-|X|}\, \int \prod_{x\in X} d\Phi (x)\, \exp \{
   -S(X|\Phi , \kappa ) + \sum_{x\in X} J(x) \Phi (x)\} ,
\end{equation}
where the action is
\[
S(X|\Phi , \kappa ) = \sum_{x\in X} V(\Phi (x)) -
  \frac{1}{2} \sum_{x,y\in X}
     \Phi (x) v(x,y) \Phi (y),
\]
and $|X|$ denotes the number of lattice sites of $X$.
The following lemma states that the partition functions are well-defined
if the local interaction $V$ obeys the {\em stability bound}
(\ref{Stab}).
\begin{lemma} \label{lem2}
Suppose that there exist positive $c>0$ and real $\delta $
such that the stability bound for the local interaction eq. (\ref{Stab})
is valid.
Then we have for all $\kappa \in \mathbb{C} $ and $\epsilon >0$ obeying
$4d|\kappa | + \epsilon < c$, and for all subsets $X\subseteq\Lambda$
\[
S(X \vert \Phi ,\kappa ) \ge \epsilon \sum_{x\in X}
    \Phi^2(x) - \delta |X|,
\]
and the integral on the right hand side of Eq.~\eqn{e7} is convergent.
\end{lemma}
This follows from the stability bound \eqn{Stab} and Lemma \ref{lem1}.

In contrast to the convergence proof of \cite{andreas1} given
for nearest-neighbour couplings
we need here a more general notion of polymers.
Here we call {\em all} finite nonempty subsets $X$
of $\Lambda $ {\em polymers}. Define {\em Mayer activities} for
polymers $X$ by
\begin{equation} \label{e8}
M(X|J,\kappa ) := -\delta_{1,|X|} + \sum_{n\ge 1} (-1)^{n-1} (n-1)! \,
      \sum_{X=\sum_{i=1}^n Q_i} \prod_{i=1}^n Z(Q_i|J,\kappa ).
\end{equation}
Then, the Mayer Montroll equation for the connected 2-point Green function
reads (cf. \cite{MacPor89} for a proof),
for all $x,y \in \Lambda $,
\bea \label{e9}
W^{(2)}(x,y) &=&
\sum_{\rm Q:\, Q\; polymer} \; \sum_{n\ge 0}
 \sum_{\cP_1,\ldots ,\cP_n \in K(Q) \atop \emptyset \ne \cP_i \in
  Conn(\cP_{i-1})} \frac{\partial^2 M(Q |J,\kappa ) }{\partial J(x)
      \partial J(y)}
\nonumber \\ & &
     (-M)^{\cP_1+\cdots +\cP_n}\vert_{J=0}
\eea
where we used the following definitions
\[
K(X) := \{ \cP = \{ P_1,\ldots ,P_n \} |\, P_i \subseteq X,\,
  P_a \cap P_b = \emptyset , \,  \forall a \ne b,\,
  n\in \mathbb{N} \cup \{ 0\} \}
\]
and $\cP_0 := \{ \{ x\} |\, x\in Q\} $ in the sum on the right hand side
of eq. (\ref{e9}). Furthermore, for a set $\cP = \{ P_1,\ldots ,P_n\} $
consisting of polymers $P_a$ we have defined
\[
Conn (\cP ) := \{ \cP^{\prime} \in K(\Lambda ) |\, \forall P^{\prime}
  \in \cP^{\prime} \, \exists a \in \{ 1,\ldots ,n\} : \,
  P^{\prime} \cap P_a \ne \emptyset \}
\]
and used the abbreviation
\[
(-M)^{\cP_1 +\cdots +\cP_n} := \prod_{i=1}^n (-M(P_i|J,\kappa )).
\]
The following lemma presents the conditions under which the
Mayer Montroll expansion eq. (\ref{e9}) is convergent
and exponentially bounded (cf. \cite{KotPre86}).
For a proof see \cite{andreas1}.
\begin{lemma} \label{lem3}
Suppose that there exist positive constants $\alpha ,\kappa_* >0$
and a function $m(\kappa)>0$ such that
\begin{equation} \label{e10}
\sum_{P:\, y\in P \atop |P| \ge 2} |M(P|J=0,\, \kappa )| \, \exp \{
  \alpha |P| \} \le \frac{\alpha }{2}
\end{equation}
and
\bea \label{e11}
\lefteqn{
\sum_{P:\, y\in P }
|\frac{\partial^2 M(P|J,\kappa )}{\partial J(x) \partial J(y)}
   \vert_{J=0} | \, \exp \{ \alpha |P| \} \le }
\nonumber \\ & &
   e^\alpha \, q_V^{(2)} \delta_{xy} +
   \frac{\alpha }{2}\, \exp \{ - m \Vert x-y\Vert \} ,
\eea
for all $x,y\in \Lambda $ and $|\kappa | \le \kappa_*,$
where
\be
q_V^{(2)} := \frac{\int d^N\Phi \, \Phi^2 \,
  \exp \{ -V(\Phi )\}}{\int d^N\Phi \, \exp \{ -V(\Phi )\}} .
\ee
Then the series expansion eq. (\ref{e9}) is convergent and
\begin{equation} \label{e11a}
|W^{(2)}(x,y)| \le  e^\alpha \,  q_V^{(2)}\delta_{xy} +
 \alpha \, \exp \{ - m \Vert x-y\Vert \} .
\end{equation}
As $\kappa\to 0$,
\[
   m(\kappa) = O(\ln{|\kappa|}).
\]
\end{lemma}
The next lemma is the core of the convergence proof.
It states sufficient conditions for the supposition
of Lemma \ref{lem3} to hold.
\begin{lemma} \label{lem4}
Suppose that the stability bound (\ref{Stab}) and the supposition
of lemma \ref{lem1} are valid.
There exist positive constants $\alpha >0$ and $\kappa_* >0$
and a positive function $m(\kappa)>0$ such that
inequalities (\ref{e10}) and (\ref{e11}) are valid for all $|\kappa | \le
\kappa_* .$ Furthermore, for $\kappa \rightarrow 0,$
\[
m(\kappa ) = O(\ln |\kappa |).
\]
\end{lemma}
\par\noindent
{\em Proof :} For the proof we will use the tree graph formula
\cite{AbdRiv94} for Mayer activities. Let $X$ be a polymer, $|X| \ge 2.$
Then
\bea \label{Tree}
\lefteqn{M(X|J,\kappa ) = }
\nonumber\\ & &
\sum_{\tau :\, \tau \in T(X)}
\int \prod_{x\in X} d\Phi (x) \,
\exp \{ \sum_{x\in X}(-V(\Phi (x)) +
  J(x) \Phi (x) )\}\, [\prod_{(xy)\in \tau} \int_0^1 dt_{xy}]
\nonumber\\ & &
\prod_{<xy> \in \tau} (\Phi (x) v(x,y) \Phi (y)) \,
\exp \{ \frac{1}{2}\sum_{x,y\in X,\ x\ne y} t_\tau^{\mbox{min}}(x,y) \,
  \Phi (x)v(x,y) \Phi (y) \} .
\eea
In the following we will explain the notations used in formula
(\ref{Tree}). $T(X)$ denotes the set of all tree graphs (graph
containing no loops) with lines $(xy),$ $x\in X$ and vertex set $X.$
Furthermore,
\bela
\lefteqn{t_\tau^{\mbox{min}}(x,y) := }
\nonumber\\ & &
\mbox{min}\, \{ t_l|\, \mbox{ path
   connecting $x$ and $y$ and containing link $l\in \tau $} \} .
\eela
Using the stability bound (\ref{Stab}),
definition (\ref{qdef}) of the constant $q_V,$
Lemma \ref{lem2} and the tree graph formula (\ref{Tree}) we obtain
\bea \label{e22}
|M(X|J=0,\kappa )| &\le & \sum_{\tau :\, \tau \in T(X)}
 \left( \prod_{(xy)\in \tau}
 |v(x,y)|  \right) q_V^{-|X|} e^{\delta |X|}
\nonumber \\ & &
\epsilon^{-\frac{\sum_{x\in X} (d_\tau (x) +1)}{2}} \, \prod_{x\in X}
(\frac{d_\tau (x) + 1}{2})!
\eea
where
\[
d_\tau(x) := |\{ l\in \tau |\, \mbox{line $l$ emerges from vertex $x$} \} |
\]
is the number of lines in the tree graph $\tau$ which are connected
to vertex $x.$ For the estimation of the fields in the term
$\prod_{<xy>\in \tau} \Phi (x) v(x,y) \Phi (y) $ we have used
the stability bound (\ref{Stab}) and
\[
2\int_0^\infty dx\, \exp \{ -\epsilon x^2 \} x^n =
  \epsilon^{-\frac{n+1}{2}} (\frac{n+1}{2})! ,
\]
and we write $z!\equiv\Gamma(z+1)$ for $z\geq 0$.
For $X=\{ x_1, \ldots ,x_n\} $ and $d_i \in \mathbb{N} ,$ $i\in \{ 1, \ldots ,
n\} $ let $T(X; d_1,\ldots , d_n)$ be the set of all tree graphs $\tau \in
T(X)$ such that $d_\tau (x_i) = d_i$ for all $i\in \{ 1, \ldots ,
n\} .$ Cayley's Theorem ( for a proof cf. \cite{Har73}) is
\begin{equation} \label{e23}
|T(X; d_1,\ldots , d_n)| = \frac{(n-2)!}{(d_1-1)! \cdots (d_n-1)!} .
\end{equation}
In eq. (\ref{e22}) we replace the sum over tree graphs by
\[
\sum_{\tau :\, \tau \in T(X)} = \sum_{d_1, \ldots ,d_n \ge 1 \atop
 \sum_{i=1}^n d_i = 2(n-1)} \sum_{\tau \in T(X; d_1,\ldots , d_n)} .
\]
Using that
\[
(\frac{d+1}{2})! \le (d-1)!
\]
for all $d\ge 3$ and that there exists at least one $x\in X$ such that
$d_\tau (x) =1$, we obtain
\begin{equation} \label{e25}
\prod_{x\in X} (\frac{d_\tau (x) + 1}{2})! \le
(\frac{3}{2})!^{|X|-1} \, \prod_{x\in X} (d_\tau (x) -1)!.
\end{equation}
Furthermore,
\bea \label{e26}
\lefteqn{\sum_{X:\, y\in X\atop |X| \ge 2} |M(X|J=0,\kappa ) |\,
  \exp \{ \alpha |X| \} }
\nonumber \\  &=&
 \sum_{n:\, n \ge 2} \frac{1}{(n-1)!} \sum_{x_2,\ldots ,x_n\atop
   x_a \ne x_b, \, a\ne b} |M(\{ y,x_2,\ldots ,x_n\} | J=0,\kappa )|
   \exp \{ \alpha n \} .
\eea
We have replaced here the sum over all polymers $X$ with $|X| \ge 2,$
$y\in X,$ by a sum over distinct labelled lattice sites
$x_2,\ldots ,x_n.$ We have defined $x_1 := y.$ Then a tree $\tau \in
T(\{ x_1,\ldots ,x_n\} )$ corresponds to a tree $T(\{ 1,\ldots ,n\} )$
where the vertices are labelled by $1,\ldots ,n.$
We want to estimate the sum over all lattice sites $x_2,\ldots ,x_n$
for a given tree $\tau .$ For that we have to find a bound for
\begin{equation} \label{e27}
\sum_{x_2,\ldots ,x_n\atop x_a \ne x_b, \, a\ne b}
  \prod_{(ab)\in \tau } |v(x_a,x_b)|
\end{equation}
for all tree graphs $\tau \in T(\{ 1, \ldots ,n\} ).$
This can be done by estimating
\[
\sum_{x^{\prime} :\, x^{\prime} \in \Lambda} |v(x,x^{\prime})|.
\]
To obtain later in this proof the exponential factor
$\exp \{ -m\Vert x-y\Vert \} $ in the bound (\ref{e11})
we find an upper bound for
\[
\sum_{x^{\prime} :\, x^{\prime} \in \Lambda} |v(x,x^{\prime})|
(\frac{\kappa_*}{|\kappa |})^{\Vert x-x^{\prime}\Vert }.
\]
Using the suppositions (\ref{e2}) and (\ref{e3}) and \eqn{res.6},
we see that for $\kappa$ with $|\kappa|\leq \kappa_* \leq \overline\kappa$
\bea \label{e28}
\sum_{x^{\prime} :\, x^{\prime} \in \Lambda} |v(x,x^{\prime})|
  (\frac{\kappa_*}{|\kappa |})^{\Vert x-x^{\prime}\Vert }
    & \le &
 \sum_{x^{\prime} :\,x\not= x^{\prime} \in \Lambda}
  \; \sum_{p:\, l_p = \Vert x-x^{\prime} \Vert } 2 |c_p| \kappa_*^{l_p}
\nonumber \\  & \le &
  2 \;\sum_{x^{\prime} :\, x\not= x^{\prime} \in \Lambda}
   \; \sum_{p:\, l_p = \Vert x-x^{\prime}\Vert } 2^{-l_p} 2^{-\eta \Vert
    x - x^{\prime} \Vert }
   \left(\frac{\kappa_*}{\overline\kappa}\right)^{l_p} \overline\kappa
\nonumber \\  & \le &
8 D \kappa_*.
\eea
We have used that $l_p\geq 1$ for all signatures $p$.
We sum in expression (\ref{e27}) successively over all
$x_2,\ldots ,x_n \in \Lambda $ starting with vertices having
a maximal distance to the ``root'' vertex $y=x_1.$ This gives
a bound
\begin{equation} \label{e28a}
\sum_{x_2,\ldots ,x_n\atop x_a \ne x_b, \, a\ne b}
  \prod_{(ab)\in \tau } |v(x_a,x_b)|
  \; \leq \; (8D|\kappa|)^{n-1}.
\end{equation}
Thus, using
\[
\sum_{d_1,\ldots ,d_n \ge 1 \atop \sum_{i=1}^n d_i = 2(n-1)} 1 \le
  2^{2(n-1)},
\]
Cayley's theorem Eq.~(\ref{e23}), inequalities (\ref{e25}) and (\ref{e26})
and the bound (\ref{e28a}), we obtain
\bea \label{e29}
\lefteqn{\sum_{X:\, y\in X\atop |X| \ge 2} |M(X|J=0,\kappa ) |\,
  \exp \{ \alpha |X| \} }
\nonumber \\  &\le &
 \sum_{n:\, n \ge 2} \frac{1}{n-1}\,
  (32 (\frac{3}{2})!\,D \kappa_* )^{n-1} q_V^{-n}
  \epsilon^{-\frac{3n-2}{2}}  \exp \{ (\alpha + \delta )n \}
\nonumber\\ &=&
\sum_{n\ge 1} \frac{1}{n} (32 (\frac{3}{2})!\,
D \kappa_* q_V^{-1} \epsilon^{-\frac{3}{2}} e^{\alpha +\delta })^n
 e^{\alpha+\delta} q_V^{-1} \epsilon^{-\frac{1}{2}} =
  -\ln (1-u)\, e^{\alpha +\delta } q_V^{-1} \epsilon^{-\frac{1}{2}} ,
\eea
where
\[
u := 32 (\frac{3}{2})!\, D \kappa_* q_V^{-1}
     \epsilon^{-\frac{3}{2}}e^{\alpha +\delta }.
\]
The proof of the bound of the term on the left hand side of (\ref{e11})
is similar to the proof of inequality (\ref{e29}). To obtain the
exponential factor $\exp \{ -m\Vert x-y\Vert \} $ we use the fact that
in the trees $\tau$ occuring in tree graph formula of
$\frac{\partial^2 M(X|J,\kappa )}{\partial J(x) \partial J(y)} $
there exists a path connecting the vertices $x$ and $y.$ Then we use the
bound
\[
 \prod_{(ab)\in \tau } |v(x_a,x_b)| \le (\frac{|\kappa|}{\kappa_*})^{\Vert
   x-y\Vert }  \prod_{(ab)\in \tau } |v(x_a,x_b)| \,
  (\frac{\kappa_*}{|\kappa|})^{\Vert x_a-x_b\Vert } .
\]
For the sum over all positions $x_2,\ldots ,x_n$ for a given tree $\tau$
we use inequality (\ref{e28}). The result is
\bea \label{e30}
\lefteqn{\sum_{X:\, y\in X\atop |X| \ge 2} |
\frac{\partial^2 M(X|J,\kappa )}{\partial J(x) \partial J(y)} \vert_{J=0} |\,
  \exp \{ \alpha |X| \} }
\nonumber \\  &\le &
  -\ln (1-u)\, e^{\alpha +\delta } \epsilon^{-\frac{3}{2}}
    \, \exp \{ -m\Vert x-y\Vert \} ,
\eea
where
\[
m = \ln (\frac{\kappa_*}{|\kappa |}) .
\]
This proves the assertion if $\kappa_* $
is small enough. $\qquad \square $

Lemmata \ref{lem3} and \ref{lem4} prove that the Mayer Montroll equations
(\ref{e9}) are absolutely convergent under the suppositions of
theorem \ref{the1}. By a standard theorem for analytic functions we
see that theorem \ref{the1} holds.

\section{Summary}
We have discussed the linked cluster expansion in application to lattice
models with general pair interactions beyond nearest neighbour
couplings. Under very general conditions, free energy, truncated
correlation functions and susceptibilities are uniformly
convergent in volume to analytic functions,
for sufficiently small hopping parameters.

The graphical expansion is cast in terms of graph classes
well known from the pure nearest neighbour interactions on
hypercubic lattices. This implies that
all the simplifications and computational efforts
that are based on the topology of those lattices
are fully exploited.
The only modification occurs as a generalization of free random walks.

\section*{Acknowledgement}
A.P. would like to thank the
Deutsche For\-schungs\-ge\-mein\-schaft for financial support.

\begin{appendix}
\end{appendix}
%
%


\end{document}